\documentclass[useAMS,usenatbib]{mn2e}

\usepackage{graphicx}
\usepackage{mathrsfs}
\usepackage{natbib}
\usepackage{amssymb}
\usepackage{amsmath}
\usepackage{mathtools}
\usepackage{overpic}
\bibpunct[,]{(}{)}{;}{a}{}{,}

\def\totd{{\mathrm{d}}}
\def\sun{{\odot}}

\title[Effect of BH spin on outflows from NS merger remnant disks]{Outflows from accretion disks formed in neutron star mergers:
effect of black hole spin}
\author[Fern\'andez, Kasen, Metzger, \& Quataert]{Rodrigo Fern\'andez$^{1,2}$, 
Daniel Kasen$^{1,3}$, Brian~D. Metzger$^4$, Eliot Quataert$^2$\\
$^1$ Department of Physics, University of California, Berkeley, CA 94720, USA.\\
$^2$ Department of Astronomy \& Theoretical Astrophysics Center, University of California, Berkeley, CA 94720, USA.\\
$^3$ Nuclear Science Division, Lawrence Berkeley National Laboratory, Berkeley, CA 94720, USA.\\
$^4$ Columbia Astrophysics Laboratory, Columbia University, New York, NY 10027, USA.
}

\begin{document}

\date{Submitted to MNRAS}
\pagerange{\pageref{firstpage}--\pageref{lastpage}} 
\pubyear{2014}
\maketitle
\label{firstpage}

\begin{abstract}
The accretion disk that forms after a neutron star merger
is a source of neutron-rich ejecta. The ejected material contributes to a 
radioactively-powered electromagnetic transient, with properties
that depend sensitively on the composition of the outflow.  
Here we investigate how the spin of the black hole
remnant influences mass ejection on the thermal and viscous timescales. 
We carry out two-dimensional, time-dependent
hydrodynamic simulations of merger remnant accretion disks including
viscous angular momentum transport and approximate
neutrino self-irradiation. The gravity of the spinning black hole is
included via a pseudo-Newtonian potential. We find that 
a disk around a spinning black hole ejects more mass, up to a factor of several,
relative to the non-spinning case. The enhanced mass loss is due to
energy release by accretion occurring deeper in the gravitational potential, raising
the disk temperature and hence the rate of viscous heating in regions where
neutrino cooling is ineffective.
The mean electron fraction of the outflow increases moderately with BH spin
due to a highly-irradiated (though not neutrino-driven) wind component.
While the bulk of the ejecta is still very neutron-rich,  thus 
generating heavy $r$-process elements, the leading edge of the wind contains a small 
amount of Lanthanide-free material. This component can give rise 
to a $\lesssim 1$~day blue optical `bump' in a kilonova light curve, even in the case
of prompt BH formation, which may facilitate its detection.
\end{abstract}


\maketitle

\section{Introduction}

Binary neutron star (NS) and neutron star -- black hole (BH) mergers are prime candidates for 
detection of gravitational waves (GW) with upcoming ground-based interferometers such
as Advanced LIGO, Virgo, and KAGRA (e.g., \citealt{Abadie+10}). Material
ejected in these mergers is expected to generate an electromagnetic (EM)
counterpart powered by the radioactive decay of elements synthesized
during the expansion \citep{Li&Paczynski98,Metzger+10b,Roberts+11,Grossman+13}. 
Good knowledge of the properties of these EM
counterparts would help with planning follow-up searches of GW sources 
(e.g., \citealt{Bloom+09,Metzger&Berger12,Nissanke+13,ghosh2014,kasliwal2014,singer2014}).

The nature of the EM counterpart depends crucially on the optical opacity of the 
ejected material, which is sensitive to composition.
In particular, very neutron-rich material (with electron fractions 
$Y_e \lesssim 0.2$) produces heavy $r$-process elements upon decompression, 
a prediction that allows these sources to be a leading candidate for the $r$-process site 
\citep{Lattimer&Schramm74,Freiburghaus+99,Korobkin+12,goriely2013}. 
The presence of elements with mass number 
$A>140$ is also accompanied by a dramatic increase in the optical opacity 
relative to iron-peak matter \citep{Kasen+13}. The resulting EM supernova-like
counterpart peaks in the near infrared (IR) band on a timescale of 
$\sim$~weeks \citep{Barnes&Kasen13,tanaka2013}.
Such an $r$-process \emph{kilonova} may have already been detected
in the afterglow of GRB 130603B \citep{Tanvir+13,Berger+13}.

Mergers can eject material via three mechanisms. First, tidal forces
lead to the expulsion of matter on the dynamical 
time (e.g., \citealt{Korobkin+12,bauswein2013,Hotokezaka+13,foucart2014}).
The amount of material ejected depends on the mass ratio and the spin of
both components, although in most cases this \emph{dynamical ejecta} is very neutron-rich
($Y_e\lesssim 0.1$), has high velocities ($\sim 0.1-0.3$c), and is 
confined mostly to the orbital plane. 
Second, neutrino irradiation of 
the remnant accretion disk can lead to a thermally-driven 
wind (e.g., \citealt{McLaughlin&Surman05,surman2006,Surman+08,Wanajo&Janka12}). 
Whether this mechanism leads to any mass ejection is highly
dependent on the level of neutrino irradiation, becoming most important
when a hypermassive neutron star (HMNS) phase precedes BH formation
(e.g., \citealt{Dessart+09,MF14,perego2014}). This outflow is
generally driven toward polar regions, carrying a higher proton
content than the dynamical ejecta ($Y_e\sim 0.3-0.4$).

Third, outflows are driven on longer timescales by
a combination of angular momentum transport, nuclear recombination,
and the decrease in neutrino emission as the disk temperature 
falls \citep{Metzger+09a,Lee+09}. When the central object is a
promptly-formed, non-spinning BH, this mechanism can eject several 
percent of the initial disk mass in a neutron-rich and 
quasi-spherical outflow (\citealt{FM13}, hereafter FM13). A long-lived HMNS leads to 
significantly larger mass ejection, with higher mean electron fraction
($Y_e\sim 0.3-0.4$).
Recently, \citet{just2014} have examined this mass ejection mechanism
including the effect of BH spin and employing
energy-dependent neutrino transport, finding
that a larger fraction of the initial disk mass ($\sim 20\%$)
is ejected, with higher electron fractions ($Y_e\sim 0.3$), relative
to the non-spinning case studied by FM13 and \citet{MF14}.

Numerical simulations of NS-NS mergers find remnant black hole
spins in the range $a = 0.7-0.9$, with the exact value depending
on the mass ratio of the binary, the spin of each component, the
equation of state, and the numerical resolution \citep{oechslin2007,Kiuchi+09,Rezzolla+10}.
In addition, general relativistic magnetohydrodynamic
simulations of radiatively-inefficient accretion disks around BHs 
find that the magnitude of the energy and mass ejection in the 
non-relativistic wind depend on the spin of the BH \citep{mckinney2012,sadowski2013}
 
The goal of this paper is to systematically study the effect
of the gravitational potential of a spinning BH on the magnitude
and composition of the outflows from remnant accretion disks. 
To this end, we extend the approach of FM13 and \citet{MF14} to include a
pseudo-Newtonian potential that models the
spacetime of a spinning BH. 
We employ two different prescriptions for neutrino
irradiation in order to separate differences in the 
results due to gravity and weak interactions.
While we do not include general relativity in our simulations, 
the effects of a smaller innermost stable orbit on the outflow
obtained with a pseudo-Newtonian potential should 
persist when adding more detailed physics, since basic
considerations dictate that more energy is released around
a spinning BH.

The structure of this paper is the following. Section \ref{s:methods}
summarizes the numerical implementation, discusses
the properties of the pseudo-Newtonian potential, and describes
the list of models evolved. Results are presented in \S\ref{s:results},
separated into dynamics and composition of the outflow.
Our main conclusions are summarized in \S\ref{s:summary}, where
we also discuss implications for EM counterparts of GW sources,
and the production of $r$-process elements.

\section{Methods}
\label{s:methods}

\subsection{Equations, Microphysics, and Numerical Approach}

We solve the compressible hydrodynamic equations using 
FLASH3.2 \citep{dubey2009}. The
public version of the code has been modified to include a non-uniform
grid \citep{F12} and source terms to model the long-term evolution
of compact object merger remnant accretion disks (FM13; \citealt{FM12,MF14}).
The equation of state is that of \citet{timmes2000}, with abundances
of neutrons, protons and alpha particles satisfying nuclear statistical
equilibrium (NSE) above a temperature of $5\times 10^9$~K, and
accounting for the nuclear binding energy of alpha particles.

Angular momentum transport is mediated by an anomalous shear stress
tensor, using the kinematic viscosity of \citet{shakura1973}. The gravity
of the BH is modeled by a pseudo-Newtonian potential (\S\ref{s:gravity}),
ignoring the contribution from the disk.

Neutrino source terms are modeled in two ways.
The first implementation is that of FM13, which includes a geometric self-irradiation
prescription that applies finite optical depth corrections to otherwise 
optically thin emissivities. This method is suitable for disks that are
at most marginally optically thick to neutrinos. The second approach is
a neutrino leakage scheme \citep{MF14} based on the method of \citet{Ruffert+97}.
In both cases, the evolution of the electron fraction and the net
energy deposition are followed self-consistently. 
Only charged-current weak interactions are included.

The computational domain is discretized in spherical polar coordinates,
with a radial grid logarithmically spaced and an angular grid equispaced
in $\cos\theta$, covering the interval $\theta \in [0,\pi]$. The inner
radial boundary is placed midway between the innermost stable
circular orbit (ISCO) and the horizon radius. The outer radial boundary
is located at a radius $10^3$ times larger. A total of $192$ cells
in radius and $56$ in angle are used, corresponding to
$\Delta r /r \simeq \Delta \theta \simeq 2^\circ$ at the equator.
One model is evolved at double resolution in angle and radius to
verify convergence.
The boundary conditions are reflecting in angle, and outflow
in radius. 

The initial condition is an equilibrium torus constructed with the Helmholtz 
EOS, nuclear abundances in NSE, the pseudo-Newtonian potential of \S\ref{s:gravity}, as well 
as constant angular momentum, entropy, and electron fraction. As in FM13 and \citet{MF14}, 
the torus is allowed to relax for $100$ orbits (measured at the initial density peak) without neutrino 
or viscous source terms, in order to separate initial transients from genuine 
disk winds. After this time, the radial velocity is reset and source terms
are turned on. We set a density floor at $1$~g~cm$^{-3}$ and a cutoff density for
source terms at $10$~g~cm$^{-3}$ to avoid numerical problems near the inner
boundary. In the case of the models with leakage scheme, this floor has a 
radial variation $\propto r^{-3}$, starting at values $10$ times higher 
than the fiducial constant values at the inner boundary.

\begin{table*}
\centering
\begin{minipage}{16.5cm}
\caption{List of evolved models and summary of results\label{t:models}\label{t:results}. The first
five columns show model name, BH spin parameter, ISCO radius (in units of $r_g = GM_{\rm bh}/c^2$), 
initial torus mass, and 
type of neutrino treatment (Leak: leakage scheme, Thin: optically-thin with corrections), respectively.
The following six columns show integrated properties of the outflow, restricted to equatorial
 ($\theta \in [30^\circ,150^\circ]$) and polar ($\theta<30^\circ$ and $\theta > 150^\circ$)
latitudes. The three columns in each group show the mass-flux weighted (eq.~\ref{eq:mass_flux_average}) 
electron fraction, entropy,
and expansion time at a radius where the mass-flux weighted temperature is $5\times10^9$~K,
respectively. The 12th column shows the ratio of polar to equatorial mass ejection at
the radius where the polar outflow is measured, when present. The last two columns show the 
mass ejection at all angles, total and unbound, at a radius of $\sim 10^9$~cm in units of 
the initial torus mass.}
\begin{tabular}{lccccccccccccc}
\hline
{Model}&
{$a$} &
{$r_{\rm ISCO}$} &
{$M_{\rm t0}$} &
{Neutrinos} &
\multicolumn{3}{c}{Equatorial Outflow} &
\multicolumn{3}{c}{Polar Outflow$^b$} &
{} & {} & {}\\
{ } & {} & {} & {} & {} &
{$\bar{Y}_e$} &
{$\bar{s}$ } &
{$t_{\rm exp}$} &
{$\bar{Y}_e$} &
{$\bar{s}$ } &
{$t_{\rm exp}$} &
{$M_{\rm pol}$}    &
{$M_{\rm ej,tot}$} &
{$M_{\rm ej,unb}$} \\
\multicolumn{2}{c}{} & {($r_{\rm g}$)} & {($M_\odot$)} & {} &
{} & {($k$/b)} & {(ms)} & {} & {($k$/b)} & {(ms)} & {$(M_{\rm eq})$} & {($M_{\rm t0}$)} & {($M_{\rm t0}$)} \\
\hline
t-a00      & 0.00   & 6.00 & 0.03 & Leak  & 0.19 & 19 & 31 & ...  & ... & ... & ...  & 0.05 & 0.03 \\ 
t-a40      & 0.40   & 4.61 &      &       & 0.21 & 21 & 37 & ...  & ... & ... & ...  & 0.09 & 0.06 \\ 
t-a80      & 0.80   & 2.91 &      &       & 0.22 & 21 & 35 & 0.31 & 38  & 9.4 & 0.01 & 0.17 & 0.11 \\ 
t-a95      & 0.95   & 1.94 &      &       & 0.24 & 21 & 42 & 0.32 & 51  &  10 & 0.03 & 0.28 & 0.21 \\ 
\noalign{\smallskip}                                                                      
s-a00      & 0.00   & 6.00 & 0.03 & Thin  & 0.18 & 20 & 35 & ...  & ... & ... & ...  & 0.05 & 0.03 \\ 
s-a40      & 0.40   & 4.61 &      &       & 0.20 & 21 & 38 & ...  & ... & ... & ...  & 0.10 & 0.07 \\ 
s-a80      & 0.80   & 2.91 &      &       & 0.22 & 21 & 38 & ...  & ... & ... & ...  & 0.19 & 0.13 \\ 
s-a95      & 0.95   & 1.94 &      &       & 0.24 & 22 & 34 & 0.32 & 56  & 11  & 0.02 & 0.30 & 0.22 \\ 
\noalign{\smallskip}                                                                  
t-a80-hr   & 0.80   & 2.91 & 0.03 & Leak  & 0.22 & 22 & 31 & 0.31 & 45  & 10  & 0.02 & 0.19 & 0.12 \\ 
t-a00-bnd  & 0.00   & 6.00 &      &       & 0.18 & 19 & 35 & ...  & ... & ... & ...  & 0.05 & 0.03 \\ 
\noalign{\smallskip}
j-a80v2    & 0.80   & 2.91 & 0.30 & Leak  & 0.23 & 15 & 95 & 0.38 & 53  &  17 & 0.02 & 0.09 & 0.06 \\ 
j-a80v5    &        &      & 0.30 &       & 0.20 & 17 & 54 & 0.29 & 28  &  16 & 0.04 & 0.17 & 0.09 \\ 
j-a80m1    &        &      & 0.10 &       & 0.24 & 17 & 59 & ...  & ... & ... & ...  & 0.12 & 0.08 \\ 
\hline
\hline
\label{table:models}
\end{tabular}
\end{minipage}
\end{table*}

\subsection{Gravity of a Spinning BH}
\label{s:gravity}

We account for the gravity of a spinning BH through the pseudo-Newtonian
potential of \citet{artemova1996}. This potential reproduces the positions
of the ISCO and horizon of the Kerr metric, and yields steady-state, sub-Eddington, 
thin accretion disk models that differ $10\%-20\%$ from the exact general-relativistic
solution as computed by \citet{artemova1996}. In the context of time-dependent NS merger remnant accretion disk
models, this potential has been employed by \citet{setiawan2004}, \citet{setiawan2006},
and \citet{just2014}.

\citet{artemova1996} provide an analytic expression for the gravitational
acceleration
\begin{equation}
\label{eq:g_artemova}
\mathbf{g}_{\rm A} = -\nabla \Phi_{\rm A} = -\frac{GM_{\rm bh}}{r^{2-\beta}(r-r_{\rm H})^{\beta}}\hat r,
\end{equation}
where $\Phi_{\rm A}$ is the potential, $G$ is the gravitational constant, $M_{\rm bh}$
is the BH mass, $r_{\rm H}$ is the horizon radius of the Kerr metric
\begin{equation}
\label{eq:r_horizon}
r_{\rm H} = \left[1 + \sqrt{1-a^2}\right]r_g,
\end{equation}
with $a$ the BH spin parameter, $r_g = GM_{\rm bh}/c^2$, and $\beta$ relates the horizon
radius to the ISCO radius
\begin{equation}
\label{eq:beta_exponent}
\beta = \frac{r_{\rm ISCO}}{r_{\rm H}}-1.
\end{equation}
We use the analytic expression for $r_{\rm ISCO}(M_{\rm bh},a)$ in the Kerr
metric (e.g., \citealt{bardeen1972}).

Equation~(\ref{eq:g_artemova}) can be integrated analytically, subject
to the boundary condition $\Phi_{\rm A}\to 0$ for $r\to \infty$,
yielding
\begin{equation}
\label{eq:artemova_potential}
\Phi_{\rm A}(r) = 
\begin{dcases}
\frac{GM_{\rm bh}}{(\beta-1) r_{\rm H}}\left[1 - \left(\frac{r}{r-r_{\rm H}} \right)^{\beta-1} \right]
& \qquad (\beta\ne 1)\\
\noalign{\smallskip}
\frac{GM_{\rm bh}}{r_{\rm H}}\ln\left(1- \frac{r_{\rm H}}{r}\right)
& \qquad (\beta=1)
\end{dcases}
\end{equation}
for $r > r_{\rm H}$. This expression reduces to the potential of \citet{paczynsky1980}
when $a = 0$ ($\beta = 2$). The case $\beta=1$ is obtained for a spin
parameter $a \simeq 0.68654$. The expression for $\beta\neq 1$ behaves well
even for a spin parameter within $10^{-5}$ of this critical value: the
transition is continuous in $\beta$. For $r \gg r_{\rm H}$, the potential
asymptotes to the Newtonian expression.

One can use equation~(\ref{eq:artemova_potential}) to construct equilibrium tori for
initial conditions using an arbitrary equation of state (FM13). When keeping all other parameters
constant, an increase in BH spin increases the entropy of the 
torus at fixed distortion parameter (see, e.g., FM13 for a definition
of the torus distortion parameter $d$). Imposing a fixed 
entropy leads to changes of $\sim 10\%$ in the distortion parameter when going 
from no-spin to near maximal spin. All other changes in the torus are minor.

While improved potential models exist relative to that of \citet{paczynsky1980}
for non-spinning BHs (e.g., \citealt{tejeda2013}), to our knowledge no such
improvements have been developed for the spinning case.

\subsection{Models Evolved}
\label{s:models}

We evolve two sequences of models with the spin rate of the BH as the
only free parameter, as shown in Table~\ref{t:models}. 
The baseline parameter set corresponds to the
fiducial models S-def of FM13 and t000A3 (prompt BH) of
\citet{MF14}: a BH mass $M_{\rm bh} = 3M_\sun$, an initial disk
mass $M_{\rm t0}=0.03M_\sun$ with density peak located at a radial 
position $R_0 = 50$~km, uniform electron fraction and entropy
$Y_e = 0.1$ and $s = 8k_{\rm B}$/baryon, respectively, and 
a \citet{shakura1973} parameter $\alpha=0.03$. 
The ambient medium is set to have a density of $\sim 1.1$~g/cm$^3$. 

The two sequences probe the same spin rates: $a = \{0,0.4,0.8,0.95\}$,
which sample a factor of $3$ in ISCO radii, as shown in Table~\ref{t:models}.
The sequences differ in the treatment of neutrinos, in order to assess
the quantitative dispersion in the ejecta abundance and composition due to
weak processes. The first sequence (t-models) makes use of the leakage scheme described 
in \citet{MF14}, while the second (s-models) uses the semi-transparent 
approach of FM13. 

In addition, we evolve two models intended to test the consistency of
our results. The first model repeats t-a80 at double resolution in radius and angle, 
to quantify the degree of convergence of integrated wind properties.
The second model (t-a00-bnd) repeats the leakage model
with no spin (t-a00, $r_{\rm in}=4r_g$) but with an inner radial boundary located
at the same position as the model with $a=0.8$ (t-a80, $r_{\rm in} = 2.25r_g$), 
to rule out boundary effects. 

Finally, we evolve two models that mirror parameter choices made in \citet{just2014},
for comparison. Models j-a80v2 and j-a80v5 consist of a torus with
mass $M_{\rm t0} = 0.3M_\sun$, viscosity parameters $\alpha=0.02$ and $0.05$, respectively, 
and spin parameter $a = 0.8$. This combination corresponds to models
M3A8m3a2-v2 and M3A3m3a5-v2 of \citet{just2014}, which implement the same 
type of shear-stress tensor. The initial entropy of these tori is chosen to be
$6$k$_{\rm B}$ per baryon, so that the distortion parameter is comparable
to the other models ($d\simeq 2$) when going to high mass.
A third model (j-a80m1) with torus mass 
$M_{\rm t0}=0.1M_\sun$ and $\alpha=0.02$ is evolved to bridge the gap 
between these tori masses and the rest of our models.

\section{Results}
\label{s:results}

\begin{figure}
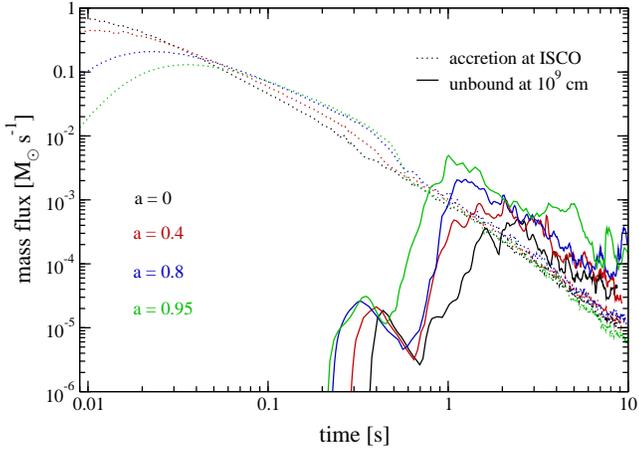

\begin{overpic}[width=\columnwidth,clip=true]{f1.eps}
\put(3.5,45.5){\tiny $\odot$}
\end{overpic}
\caption{Mass outflow rate in unbound material at $r \simeq 10^9$~cm
(solid lines) and mass accretion rate at the ISCO (dotted lines)
for models of the leakage sequence (t-series) with different BH spin.}
\label{f:mass_flux_leak}
\end{figure}

Remnant accretion disks evolve in a generic way 
\citep{popham1999,Narayan+01,Chen&Beloborodov07}. Initially,
neutrino interactions are important. Depending on the mass
of the disk and hence its optical depth, cooling of the disk 
can be delayed until the density becomes low enough
for transparency \citep{kohri2002,DiMatteo+02}. As the disk evolves further, neutrino
interactions shut down and the disk becomes radiatively-inefficient.

The bulk of mass ejection occurs after weak interactions
freeze out on a timescale of $\sim 1$~s \citep{Metzger+08b,Metzger+09a}. 
At this time, viscous
heating and nuclear recombination inject energy into the
disk, without being offset by neutrino cooling. Because weak freezout occurs
from the outside in, this outflow is driven from the outermost
parts of the disk, which have not been brought to beta
equilibrium and hence have low electron fraction.
Part of the outflow can be driven from the inner
disk at earlier times, if neutrino self-irradiation
can deposit enough energy to drive a thermal wind.

\subsection{Effect of BH spin on mass ejection}
\label{s:ejection}

\begin{figure}
\includegraphics*[width=\columnwidth]{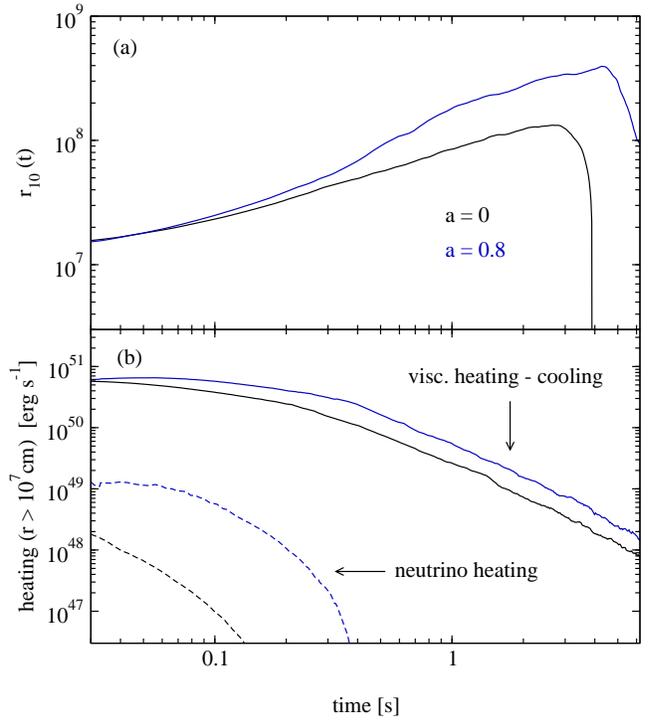}
\caption{\emph{Top}: Radius outside of which $10\%$ of the
initial disk mass ($=3\times 10^{-3}M_\odot$) resides, as a function of time, for
models t-a00 (black) and t-a80 (blue). The drop at late times occurs when
the total mass remaining in the computational domain approaches the target value.
\emph{Bottom:}
Integrated heating rate outside $r=10^7$~cm 
as a function of time for the same set of models. Solid lines show viscous
heating minus net neutrino cooling [=min$(0,Q_\nu)$], with $Q_\nu$ the net
specific neutrino energy source term, and dashed lines show net neutrino heating 
[=max$(0,Q_\nu)$].}
\label{f:heating_timescale}
\end{figure}

\begin{figure}
\includegraphics*[width=\columnwidth]{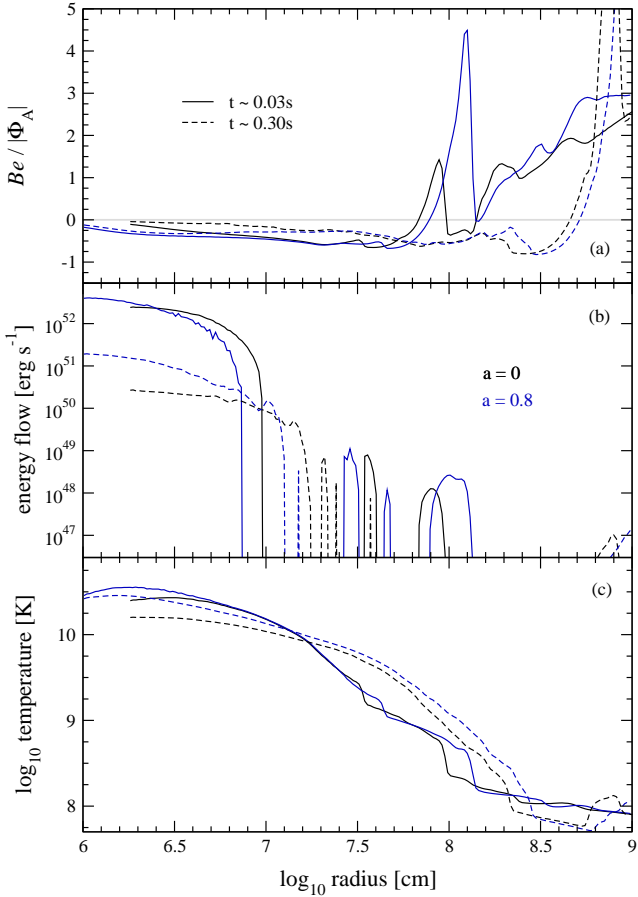}
\caption{Radial profiles at two instants in the evolution
of models t-a00 and t-a80, as labeled. Shown are (a)
the angle-averaged, density-weighted Bernoulli parameter (eq.~\ref{eq:bernoulli});
(b) the surface-integrated energy flux as a function of radius (eq.~\ref{eq:energy_flux});
and (c) the angle-averaged, density-weighted temperature. The times correspond to
a phase where neutrino cooling is important ($0.03$~s) and one in which the disk
is becoming radiatively inefficient ($0.3$~s).
Compare with Figure~\ref{f:heating_timescale}.}
\label{f:temp_radius}
\end{figure}

Mass ejection increases monotonically
with spin, as shown in Table~\ref{t:models}. In going
from $a=0$ to $a=0.95$, the amount of mass ejected
increases by a factor larger than $5$. Table~\ref{t:models}
also shows that the ratio of unbound\footnote{We define unbound
material as having positive total energy, including kinetic,
internal, and gravitational contributions.}
to bound ejecta increases slightly ($\sim 2$ to $3$) with spin.

The outflow peaks earlier for models with larger
spin, as measured\footnote{For models with $a=0.95$, the 
computational domain ends at $r=7.3\times 10^8$~cm. The wind properties
as a function of time are recorded close to this boundary and the
resulting flow is injected into a larger computational domain in order
to measure wind properties again at $r=10^9$~cm. Energy source terms are unimportant at these radii,
and are thus neglected in this procedure.} 
at a radius $\simeq 10^9$~cm from the BH
(Figure~\ref{f:mass_flux_leak}). At any
given time after models have settled into steady
accretion, the mass accretion rate through the
ISCO (which is spin-dependent) is lower for
models with higher spin. This is consistent with
existing studies of radiatively-inefficient accretion disks \citep{sadowski2013}.

To illustrate the differences in evolution introduced by a higher BH spin, 
Figure~\ref{f:heating_timescale}
shows the radius outside of which $10\%$ of the disk
mass resides, $r_{\rm 10}$, as a function of time for models t-a00 and t-a80. 
Initially, this surface evolves similarly in both models
until $\sim 0.1$~s. Afterward,
mass motion accelerates for the model with a spinning BH, settling 
later to a similar rate of expansion after $\sim 1$~s.
Figure~\ref{f:heating_timescale} also shows the net heating
rate outside $10^7$~cm
as a function of time. This heating rate 
is higher at all times 
for the model with $a=0.8$ after $\sim 30$~ms of evolution. The enhancement
is approximately a factor of $\sim 2$. 

This higher viscous energy deposition originates in the larger
amount of energy released by the accretion flow when the
inner edge of the accretion disk lies deeper in the potential well.
Figure~\ref{f:temp_radius} shows radial profiles 
at two instants in the evolution of models t-a00 and t-a80. 
Shown is the angle-averaged, density-weighted Bernoulli parameter
\begin{equation}
\label{eq:bernoulli}
{Be} = \frac{1}{2}\mathbf{v}^2 + e_{\rm int} + \frac{p}{\rho} + \Phi_{\rm A},
\end{equation}
normalized to $|\Phi_{\rm A}|$, where $\mathbf{v}$ is the total velocity 
including orbital motion, $e_{\rm int}$ is the specific internal energy including
nuclear binding energy but normalized
so that it vanishes at zero temperature, $p$ is the pressure, 
and $\rho$ is the density, as well as the surface-integrated energy flux,
\begin{equation}
\label{eq:energy_flux}
\textrm{energy flow } \equiv 2\pi\int\sin\theta\totd\theta\, r^2 \rho v_r\, Be,
\end{equation}
where the angular integral extends out to $45^\circ$ from the midplane.
The Bernoulli parameter of the accretion flow is negative in most of the disk.
When accreting ($v_r < 0$), energy thus flows outward. At early times,
when the evolution of both models is very similar and neutrino cooling is
important, the non-spinning BH
releases slightly more energy at fixed radius. When the disk starts becoming
radiatively-inefficient at $\sim 0.3$~s, however, the energy flux in the
spinning BH case is larger at all radii $r < 10^7$~cm. At this time the disk
with $a=0$ case satisfies $|Be| \ll |\Phi_{\rm A}|$ at small radii,
while the spinning BH case is more gravitationally bound at fixed radius for $r < 10^7$~cm.

The larger energy released in the spinning BH case results in a higher disk temperature 
at later times, as shown in Figure~\ref{f:temp_radius}c. 
The higher temperature increases the rate of viscous heating, given that the kinematic viscosity 
coefficient in the parameterization of \citet{shakura1973} is proportional to the sound speed squared,
with the remaining factors that enter the heating rate depending only on gradients of the
angular momentum or on position. 

A higher temperature also results in more neutrino emission, which could 
in principle compensate for a higher viscous heating rate.
However, neutrino cooling becomes unimportant for 
temperatures $T < 1$~MeV (e.g., \citealt{Metzger+09a}).
Figure~\ref{f:temp_radius} shows that this corresponds
approximately to the radius where the temperature profile
changes slope around $\sim 300$~km. 

The global contribution from neutrino heating is minor in comparison
to viscous heating.
Figure~\ref{f:heating_timescale}b shows that the integrated
neutrino heating rate at radii $r > 10^7$~cm
is only a few \% of the net viscous heating rate.
This is the case even in models with high spin, 
such as t-a80, in which there is a period
of time from about $0.03$~s to $0.3$~s during which neutrino
heating exceeds viscous heating in the radial range
$10^8\textrm{ cm} < r < 3\times 10^8$~cm. At these large radii
the magnitude of the heating is quite small, however, and
the contribution to the integrated heating rate is therefore small as well. In the
non-spinning models, neutrino heating never exceeds viscous
heating except in very localized spots at early times, with a 
negligible overall contribution.

\begin{figure}
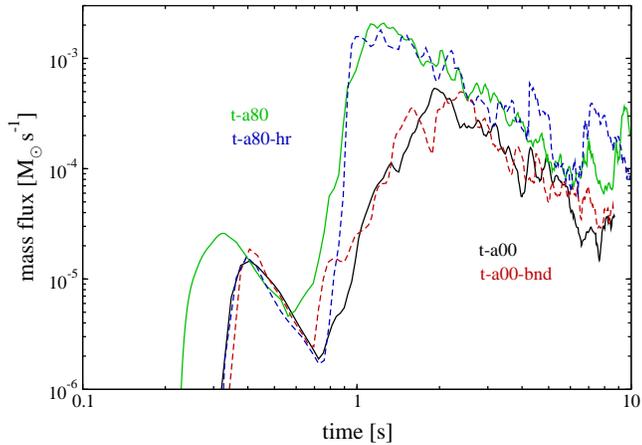

\begin{overpic}[width=\columnwidth,clip=true]{f4.eps}
\put(3.5,45.5){\tiny $\odot$}
\end{overpic}
\caption{Mass outflow rate in unbound material at $r=10^9$~cm for different models
that probe the robustness of our results. Models t-a00 (black-solid) and t-a00-bnd (red-dashed) differ only
in the location of the inner radial boundary, with the latter having a minimum
radius smaller by a factor $0.56$. Model t-a80-hr (blue-dashed) has double resolution in radius
and angle relative to model t-a80 (green-solid).}
\label{f:mout_convergence}
\end{figure}

Mass ejection in our models is essentially converged.
Table~\ref{t:models} shows that doubling the 
resolution in radius and angle leads to an overall
increase of only $\sim 10\%$ in the outflow rate, as inferred by
comparing models t-a80 and t-a80-hr. Changing the position
of the inner boundary does not modify the mass ejection
properties in any significant way, as can be seen by comparing
models t-a00 and t-a00-bnd, the latter using the
inner boundary of a spinning BH model with $a=0.8$ ($2.25r_g$ instead
of $4r_g$ for $a=0$). For comparison, the mass outflow rate at $r=10^9$~cm 
as a function of time is shown in Figure~\ref{f:mout_convergence} for the four 
models (t-a00, t-a00-bnd, t-a80, t-a80-hr).

\subsubsection{Comparison with previous work}

Existing general relativistic, magnetohydrodynamic (MHD) 
simulations of radiatively-inefficient
accretion disks find a correlation between the
amount of mass ejected in the non-relativistic wind and the
spin of the BH \citep{sadowski2013} that is qualitatively similar to what we find here. 
The character of this correlation also depends on the
initial field geometry \citep{mckinney2012}. Direct comparison of our results with these
studies can only be qualitative. On the one hand, the spatial distribution
of heating in MHD is different than that obtained with an $\alpha$ viscosity.
In addition, the difference in initial conditions leads to very different
evolutionary timescales, since these MHD studies are motivated by disks in AGN
or X-ray binaries. \citet{sadowski2013}
do not find convergence on the mass ejection from the wind,
because it is dominated by regions at large radii.
The initial tori in these simulations extend from $10r_g$ to $1000r_g$ \citep{narayan2012},
or have the initial density peak around $100r_g$ \citep{mckinney2012}. Models are evolved for a maximum
time of $2\times 10^5 r_g/c$. In contrast, our initial tori have the density
maximum at $11r_g$, and are evolved for $\simeq 6.7\times 10^5r_g/c$. A smaller
radius implies that the disk can evolve on a shorter time.

Regarding our subset of models that mirror parameters from \citet{just2014},
we find overall agreement in mass ejection for the case of model
j80-v5 at $17\%$ of the initial disk mass. This is somewhat
lower than model M3A8m3a5-v2 of \citet{just2014}, which ejects 
$23\%$ of the initial disk mass, but still consistent given the $\sim 10\%$ 
uncertainty introduced by resolution.
In contrast, our model with lower viscosity (j80-v2) ejects
only $9\%$ of the initial disk mass, whereas model M3A8m3a2-v2 of \citet{just2014}
ejects $21\%$. To ensure that this discrepant result is not
a numerical artifact, we evolved a model with $\alpha=0.02$
and $M_{\rm t0}=0.1M_\odot$, intermediate between our
fiducial model sequence and the disk masses of \citet{just2014}. 
The fraction of the disk mass ejected in this case is $12\%$, approximately
midway between what is obtained for models t-a80 and t-a80-hr
($17-19\%$) and model j80-v2. Therefore, we find a dependence
of the ejecta mass on the disk mass, with more massive disks
ejecting a lower fraction. Given that the only significant 
difference between our setup and that of \citet{just2014} is the neutrino physics, we surmise that the 
discrepant ejecta masses come from stronger cooling in our
leakage scheme at large tori masses.

\subsection{Effect of BH spin on outflow composition}
\label{s:composition}

To characterize the bulk properties of the outflow, we
take mass-flux-weighted averages as in FM13
\begin{equation}
\label{eq:mass_flux_average}
\bar{A} = \frac{\int A(r_{\rm out},\theta)\,F_{\rm M}(r_{\rm out},\theta)\,\totd\theta\totd t}
               {\int F_{\rm M}(r_{\rm out},\theta)\,\totd\theta\totd t},
\end{equation}
where $F_{\rm M} = \rho v_r$, $r_{\rm out}$ is defined so that $\bar{T} \simeq 5\times 10^9$~K, the time
integral is taken over the entire simulation, and the angular integral is restricted
to equatorial ($60^\circ$ from the equator) and polar ($30^\circ$ from the axis) latitudes.

The mean electron fraction in the equatorial 
outflow increases very modestly though monotonically with spin, as shown 
in Table~\ref{t:models}. For the neutrino leakage
sequence, $\bar{Y_e}$ increases from $0.19$ to $0.24$ when going from
$a=0$ to $a=0.95$. A similar change is obtained in the sequence
with optically thin neutrino treatment. 

\begin{figure}
\includegraphics*[width=0.5\textwidth]{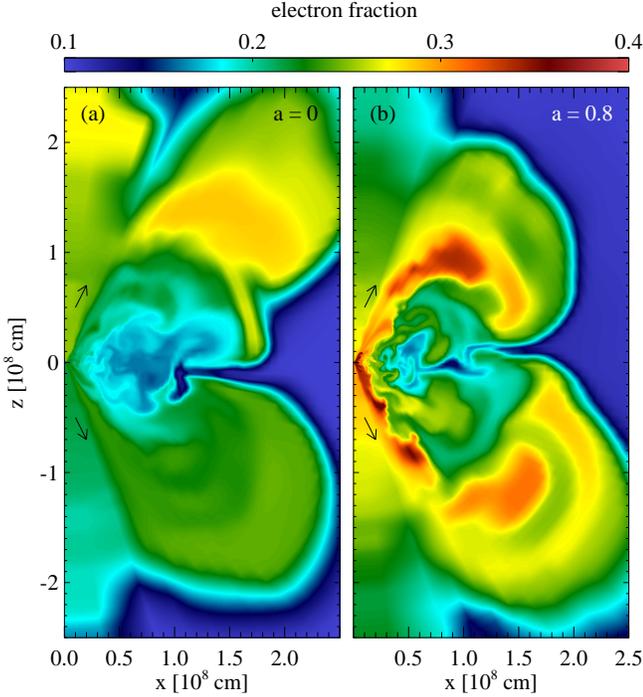}
\caption{Electron fraction in models t-a00 (left) and t-a80 (right)
at times $0.78$~s and $0.45$~s, respectively, when their winds have 
reached a similar level of expansion. The arrows indicate the
direction in which neutrino-irradiated material moves.}
\label{f:ye_launch}
\end{figure}

\begin{figure}
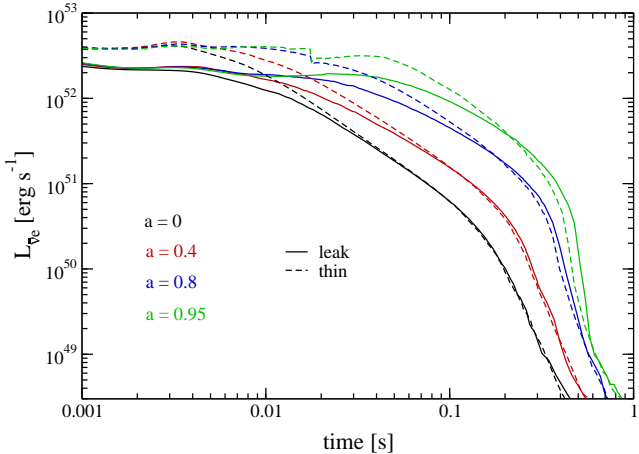

\begin{overpic}[width=\columnwidth,clip=true]{f6.eps}
\put(4,32.9){\tiny \rotatebox{90}{\_}}
\end{overpic}
\caption{Electron antineutrino luminosity for models with different
spin parameter and neutrino treatment. Shown are models in the sequence
with neutrino leakage (t-sequence, solid) and optically thin emission
with corrections (s-sequence, dashed). Colors correspond to different
BH spin parameters, as shown.}
\label{f:Lnup_spin}
\end{figure}

The origin of this increase lies in a component of the wind that is
more strongly irradiated by neutrinos than the rest of the wind,
arising from regions above the disk midplane at small radius and at
time $t\sim 0.1$~s.
Figure~\ref{f:ye_launch} illustrates how this component
differs in models with and without spin (t-a00 and t-a80). In both
cases, a radial outflow at intermediate latitudes is apparent. This
outflow wraps around the back side of the disk, and mixes with the
material in the outer disk midplane, given that the disk is convective at this time.

\begin{table}
\centering
\begin{minipage}{7cm}
\caption{Ejected mass for models in  the leakage and optically-thin
sequences, measured at $10^9$~cm and after $\sim 10$~s. Columns show from left to right model name, 
total ejected mass, ejected mass with $Y_e \geq 0.25$,
and ejected mass with $Y_e \geq 0.3$, respectively. See Table~\ref{t:models} for model parameters.}
\begin{tabular}{lccc}
\hline
{Model}&
{$M_{\rm ej,tot}$} &
{$M_{\rm ej, Y_e\geq 0.25}$} &
{$M_{\rm ej, Y_e\geq 0.3}$}\\
\noalign{\smallskip}
{} & {($\times 10^{-3}M_\odot$)} & {($M_\odot$)} & {($M_\odot$)}\\
\noalign{\smallskip}
\hline
t-a00  & 1.5 & $9.8\times 10^{-6}$ & $6.7\times 10^{-7}$ \\
t-a40  & 2.8 & $2.0\times 10^{-4}$ & $6.8\times 10^{-7}$ \\
t-a80  & 5.1 & $9.9\times 10^{-4}$ & $7.3\times 10^{-5}$ \\
t-a95  & 8.4 & $3.0\times 10^{-3}$ & $9.3\times 10^{-4}$ \\
\noalign{\smallskip}
s-a00  & 1.6 & $1.1\times 10^{-5}$ & $5.7\times 10^{-7}$  \\
s-a40  & 3.0 & $2.3\times 10^{-4}$ & $6.9\times 10^{-7}$ \\
s-a80  & 5.8 & $7.4\times 10^{-4}$ & $1.5\times 10^{-6}$ \\
s-a95  & 9.0 & $3.0\times 10^{-3}$ & $5.8\times 10^{-4}$ \\
\hline
\hline
\label{t:ejecta_composition}
\end{tabular}
\end{minipage}
\end{table}

In the spinning BH case, this irradiated component has a significantly higher
electron fraction than the rest of the disk, contributing to raise the mean
$Y_e$ of the equatorial outflow. Strong irradiation in a spinning
BH results from the neutrino luminosities remaining high
for a longer period of time, as shown in Figure~\ref{f:Lnup_spin}.
Once the neutrino luminosities decay around $t \sim 0.3$~s, 
this outflow ceases to contribute high-$Y_e$ material.
As pointed out in \S\ref{s:ejection}, neutrino heating never exceeds
viscous heating in the inner disk, hence we refer to this component
as \emph{neutrino-irradiated} instead of \emph{neutrino-driven}.

The entropy and expansion time 
at $r_{\rm out}$ in the equatorial outflow also undergo moderate
to insignificant changes with increasing BH spin, reflecting the dominance of viscous energy
deposition in setting the ouflow properties. Further support for this
hypothesis comes from the fact that overall results in the leakage and optically
thin sequences are very similar, despite non-negligible differences
in neutrino emission properties (e.g., the magnitude of the neutrino
luminosities in Figure~\ref{f:Lnup_spin}).

The signature of a neutrino-driven wind is the ejection of
material along the polar direction.
When the BH has no spin, we cannot find a value of $r_{\rm out}$
where $\bar{T}=5\times 10^9$ for polar latitudes (at this temperature
heavy elements begin to form).
This failure to find an outflow is due to the polar cavity
being mostly devoid of material. Increasing the BH spin results in more
material being ejected from the inner disk into polar latitudes.
To estimate this fraction of the outflow, we compute the total mass ejected
at $r_{\rm out}$ toward polar latitudes, whenever this quantity is
well-defined, and compare it with the total mass ejected on equatorial 
latitudes at the same radius. The result is shown in Table~\ref{t:models}. 
In all cases, the fraction of the mass ejected towards polar
latitudes is a few percent of the total outflow.
The composition of this polar outflow is noticeably different, however.
The electron fractions are $Y_e >0.3$ in all cases, with entropies
a factor of $\sim 2$ higher and expansion times a factor $\sim 3$ shorter
than the equatorial outflow.

\begin{figure*}
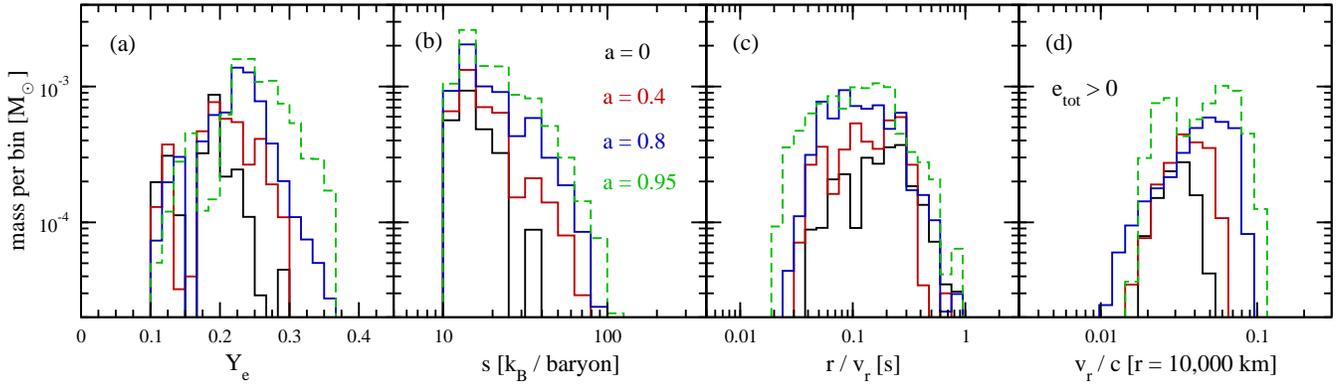

\begin{overpic}[width=\textwidth,clip=true]{f7.eps}
\put(1.3,22){\tiny $\odot$}
\end{overpic}
\caption{Mass histograms of quantities relevant for nucleosynthesis and EM transients,
for models of the t-sequence (see Table~\ref{t:models}
for model parameters). The first three panels from the left show mass distributions
as a function of electron fraction, entropy, and expansion time.
These histograms are generated 
at a radius such that the mass-flux-weighted temperature
is $\sim 5\times 10^9$~K, and including bound and unbound material. The rightmost panel
shows mass distribution as function of the velocity of the \emph{unbound} material only, computed
at $r = 10^9$~cm.}
\label{f:histogram}
\end{figure*}

The full range of nucleosynthetic- and EM transient-relevant properties in the
outflow is shown in Figure~\ref{f:histogram}, which displays mass histograms 
of electron fraction, entropy, expansion time, and asymptotic velocity
for all models in the leakage sequence. The first
three histograms from the left are constructed using mass ejected at all angles at $r_{\rm out}$, 
and include material that moves outward and inward, yielding a normalization that 
matches\footnote{FM13 only included material that moves outward.}  
the total mass ejected at $r_{\rm out}$.
The overall shape of the histograms matches the trend in 
the mass-flux-weighted quantities shown in Table~\ref{t:models}. Very similar
results are found for the optically-thin sequence.

The asymptotic velocity histogram (rightmost) shows only unbound material, and the mass
ejected is computed at $r=10^9$~cm (c.f. Figure~\ref{f:mass_flux_leak}).
The wind is sub-relativistic, with asymptotic velocities $\sim 0.05c$.
The mean velocity increases only moderately (though still monotonically) with BH spin. 

The fraction of the outflow with Lanthanide-free composition is of particular
interest to the observed kilonova properties.
Due to the lower opacities, this material should produce a bluer, optical `bump'
in the otherwise infrared kilonova, facilitating its detection at early times.
For the mean entropies and expansion times of these winds, the transition between
Lanthanide-free and Lanthanide-rich is quite abrupt, and
occurs in the range $Y_e=0.25-0.3$ (Kasen et al. 2014, in preparation).
We have computed the fraction of the total ejecta mass
with $Y_e \geq 0.25$ and $Y_e\geq 0.3$. Results are shown
in Table~\ref{t:ejecta_composition}. For the most likely
spin rate ($a=0.8$), the fraction of the ejecta that would
contribute to a `blue' precursor ranges from $1-10\%$ of the
outflow. 

\begin{figure}
\includegraphics*[width=\columnwidth]{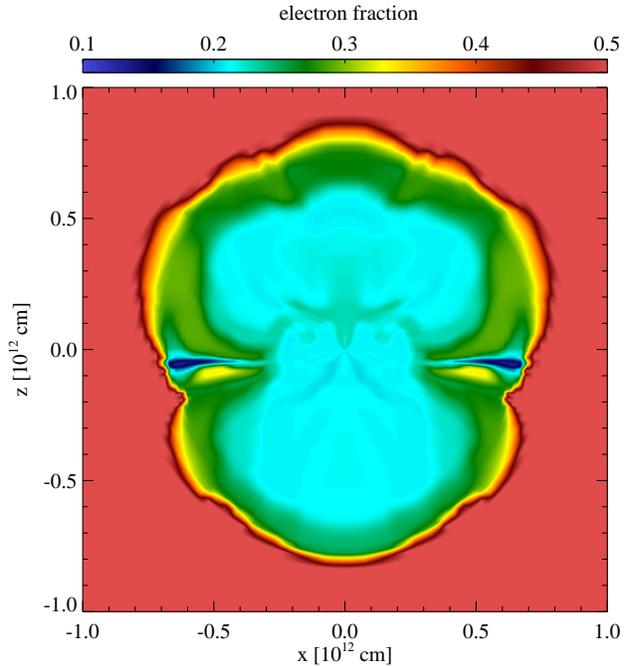}
\caption{Electron fraction for model t-a80-hr at time $t = 300$~s, when
most of the ejecta mass has reached homologous expansion. Material with
$Y_e \simeq 0.3$ and hence low optical opacity is ahead of more neutron-rich ejecta, with
the potential for the generation of a `blue' bump in the
infrared kilonova. The ambient medium has electron fraction $Y_e=1$ (pure
hydrogen, beyond the limits of the color table).}
\label{f:ye_homology}
\end{figure}

The location of the Lanthanide-free ejecta is also important, 
as it must be exterior to the high-opacity, low-$Y_e$ material,
otherwise any optical emission will be obscured.
Figure~\ref{f:ye_homology} shows the spatial
distribution of the electron fraction for model t-a80 at time $300$~s, when the velocity
distribution is very close to homologous. We obtain this
distribution by recording the properties of the wind at a radius $10^9$~cm as a function
of time. At this position, neutrino and viscous source terms operate slowly relative
to the expansion time. The sampled wind is then injected into
a much larger computational domain with lower ambient density, without energy source
terms, allowing it to expand freely for times much longer than $10$~s. A more complete description
of this procedure will be presented in Kasen et al. (2014, in preparation).
The homologous ejecta is approximately spherical, and because the higher $Y_e$
material is ejected at earlier times in the neutrino-irradiated outflow 
that wraps around the disk (Figure~\ref{f:ye_launch}),
it precedes the more neutron-rich component. This Lanthanide-free `skin' of material
would likely give rise to a blue `bump' in the kilonova light curve,
occurring on timescales of $\lesssim 1$~day and with optical luminosities
of the order of $10^{40}-10^{41}$~erg~s$^{-1}$.

\section{Summary and Discussion}
\label{s:summary}

In this paper we have systematically examined the effect of 
BH spin on the properties of
winds from NS-NS and NS-BH merger remnant accretion disks.
We have used two-dimensional, time-dependent hydrodynamic
simulations of the viscous evolution of these disks, modeling
the spacetime of a rotating BH via a pseudo-Newtonian potential,
and using two different approximations to the neutrino physics.
Our main results are the following:
\newline

\noindent
1. -- Increasing the spin of the BH increases the 
      amount of mass ejected due to weak freezout
      of the disk (Figure~\ref{f:mass_flux_leak}). 
      This enhancement can be a factor
      of $\sim 5$ for $a=0.95$ relative to the non-spinning case (Table~\ref{t:models}).
      The ratio of unbound to bound ejecta also
      increases with increasing spin.
      \newline

\noindent
2. -- The increase mass ejection at higher spin stems from
      higher net energy deposition by viscous heating (Figure~\ref{f:heating_timescale}).
      The larger heating is related to a higher energy released
      by the accretion flow given the smaller ISCO of a spinning BH
      (Figure~\ref{f:temp_radius}). Neutrino heating
      makes a minor contribution to the global energy deposition.
      This is the main reason why our two different neutrino
      implementations lead to very similar results, despite
      noticeable differences in the neutrino luminosities (Figure~\ref{f:Lnup_spin}).
      \newline

\noindent
3. -- The mean electron fraction of the outflow increases
      monotonically with increasing spin, all else being equal (Table~\ref{t:models}).
      However, the magnitude of this increase is moderate.
      For our fiducial parameter choice, increasing the 
      spin from $0$ to $0.95$ increases the mean electron
      fraction from $\sim 0.19$ to $\sim 0.24$. 
      The bulk of the outflow is still very neutron-rich.
      \newline

\noindent
4. -- The increase in electron fraction stems primarily
      from mixing of the equatorial outflow with a
      distinctive, neutrino-irradiated component. This component
      originates in regions at high altitude above the midplane and small
      cylindrical radius (Figure~\ref{f:ye_launch}). The primary driver of this
      outflow is not neutrino heating, though its composition
      is strongly affected by neutrino irradiation.
      \newline

\noindent
5. -- Other outflow properties such as the mean entropy,
      expansion time, and asymptotic velocity are not very
      sensitive to the spin of the BH (Figure~\ref{f:histogram}).
      \newline

\noindent
6. -- For the most likely spin value ($a=0.8$), the fraction
      of the ejecta that has Lanthanide-free composition
      ranges from $1$ to $10\%$ (Table~\ref{t:ejecta_composition}). 
      This high-$Y_e$ material
      resides at the leading edge of the wind (Figure~\ref{f:ye_homology}), with the
      potential to generate an early ($\lesssim 1$~day) `blue bump' in the infrared kilonova.
      \newline

Our results are in general agreement with the closely related work of \citet{just2014}.
At spins comparable with those in their models ($a=0.8$), we obtain
total outflow rates $\sim 20\%$, mean electron fractions close to 
$\sim 0.25$, and entropies $\sim 20$~k$_{\rm B}$ per baryon. While
we identify a distinctive component of the outflow that is
strongly irradiated by neutrinos, we do not find any dynamically
important region where neutrino heating dominates over viscous 
heating.
Our main difference with \citet{just2014} is quantitative, and concerns the fraction
of the ejecta mass with $Y_e$ significantly higher than $0.3$.
We attribute this discrepancy to the difference in neutrino
implementation. Given that their treatment of neutrino emission and
self-irradiation is more sophisticated than ours, our models are likely 
underestimating the fraction of the material that generates Lanthanide-free
ejecta.

While the wind from a spinning BH can in principle generate
a blue precursor to an $r$-process kilonova, there are
qualitative differences between this outflow and that
from a long-lived HMNS \citep{MF14}. First, the total amount of ejecta is
smaller by a factor of several in the BH case. Out of this material,
only a small fraction
satisfies the Lanthanide-free conditions. Second, the
geometry of the wind from a spinning BH is still mostly
spherical, while the outflow from a HMNS has a
$\sim 2:1$ asymmetry
(Kasen et al. 2014, in preparation). Therefore a long delay
to BH formation should lead to a
stronger geometric dependence and higher intensity of the `blue' precursor.

If the most likely spin of a BH remnant is $a=0.8$, then
a fraction $\sim 20\%$ of the initial disk mass
is ejected. 
Galactic chemical evolution models require an ejected
mass in $r$-process elements of $\sim 0.01M_\sun$ per event to
account for the abundance scatter in galactic halo stars
\citep{tsujimoto2014,shen2014,vdvoort2014}. 
If a prompt BH is the most likely outcome of a NS-NS or
NS-BH merger, our results suggest that these events
can produce sufficient amounts of $r$-process elements to
explain the Galactic abundance, even if the merger rate
is on the lower end of the expected value.
The absolute mass of the disk and of the dynamical ejecta can vary due to
a number of factors, and therefore the ratio of light
to heavy $r$-process elements can in principle also vary from event to
event, generating an intrinsic dispersion in the abundance pattern \citep{just2014}.

An improved estimate for the amount of mass ejected in
these winds and its composition will require the use
of general relativistic (GR) models. The highest spins
explored here ($a=0.95$) are such that the ISCO is
already close to the horizon. Our pseudo-Newtonian treatment is
thus pushed beyond its limit of validity for these
extreme cases. 

An equally important improvement concerns the inclusion
of MHD. The spatial distribution of heating is
different in a disk where MHD turbulence transports
angular momentum relative to one where viscosity
is employed (e.g., \citealt{Hirose+06}). This has implications for the
location of the wind driving and its intensity.
In addition, a disk around a spinning BH in MHD
is expected to power a jet via the Blandford-Znajek mechanism
(\citealt{blandford1977}; see also \citealt{sasha2011}), with an energy output that 
can dominate over that in the non-relativistic wind (e.g., \citealt{sadowski2013}). 
Such a jet is thought to be at the heart of the short gamma-ray burst central
engine, and its inclusion in the evolution of the system would help
to better predict the early phases of an electromagnetic transient.

Finally, a reliable prediction for the composition of the 
ejecta will require, in addition to GR and MHD, good
neutrino transport. This is essential in order to quantify
the electron fraction of the strongly-irradiated
component of the wind (whether neutrino-driven or not),
and the angle-dependent composition of the ejecta.

\section*{Acknowledgments}

We thank Oliver Just, Thomas Janka, Albino Perego, Stephan Rosswog,
Sasha Tchekhovskoy, and Francois Foucart for stimulating discussions and/or comments on the manuscript.
We also thank the referee, Maximilian Ruffert, for constructive comments that improved the paper.
RF acknowledges support from the University of California Office of the President, and
from NSF grant AST-1206097.
DK was supported in part by a Department of Energy Office of Nuclear Physics Early 
Career Award, and by the Director, Office of Energy Research, Office of High Energy and 
Nuclear Physics, Divisions of Nuclear Physics, of the U.S. Department of Energy under 
Contract No. DE-AC02-05CH11231.
BDM acknowledges support from NSF grant AST-1410950 and the Alfred P. Sloan Foundation.
EQ was supported by NSF grant AST-1206097, the David and Lucile Packard Foundation, 
and a Simons Investigator Award from the Simons Foundation.
This work was supported in part by National Science Foundation Grant No. PHYS-1066293
and the hospitality of the Aspen Center for Physics.
The software used in this work was in part developed by the DOE NNSA-ASC OASCR Flash Center at the
University of Chicago.
This research used resources of the National Energy Research Scientific Computing
Center (NERSC), which is supported by the Office of Science of the U.S. Department of Energy
under Contract No. DE-AC02-05CH11231. Computations were performed at the
\emph{Carver} cluster.

\bibliographystyle{mn2e}
\bibliography{ms}

\label{lastpage}
\end{document}